\documentclass[]{fundam}
\usepackage{url} 
\usepackage[ruled,lined]{algorithm2e}
\usepackage{graphicx}
\usepackage{tikz}
\usepackage{bussproofs}
\usepackage{multicol}
\usepackage{subcaption}
\usepackage{adjustbox}
\usetikzlibrary{automata, positioning, arrows}

\ifTUTeX
\usepackage{fontspec}
\else
\usepackage[T1]{fontenc}
\usepackage[utf8]{inputenc} 
\DeclareUnicodeCharacter{200B}{{\hskip 0pt}}
\fi
\begin{document}

%
\setcounter{page}{1}
\publyear{2021}
\papernumber{0001}
\volume{178}
\issue{1}
%

\title{Representation of a vehicular traffic model using hybrid systems}


\author{Miguel Andres Velasquez\\
Departamento de Ciencias Naturales y Matem\'aticas\\
Pontificia Universidad Javeriana-Cali \\ Calle 18 No. 118-250
Cali, Colombia.\\
miguelonvelasquez@javerianacali.edu.co
\and Carlos Ernesto Ram\'irez\\
Departamento de Ciencias Naturales y Matem\'aticas\\
Pontificia Universidad Javeriana-Cali \\ Calle 18 No. 118-250
Cali, Colombia.\\
carlosovalle@javerianacali.edu.co } 

\maketitle

\runninghead{Miguel Velasquez, Carlos Ram\'irez}{Representation of a vehicular traffic model using hybrid systems}

\begin{abstract}
 There is a great diversity of formal models to understand the dynamics of transport and vehicular flow on a road. Many of these models are inspired by the dynamics of flows governed by partial differential equations. However, it is possible to simplify these models to ordinary equations by considering constant variations in some of the input variables in this type of models. However, given that these types of systems present discrete changes when the vehicle density is altered in some sections of the lane, it seems reasonable to make use of hybrid systems to better understand the evolution of these dynamics. In this work we are interested in making use of dynamic differential logic to formally verify one of these models proposed in ordinary equations. This verification will be done through a proof assistant specially designed for hybrid systems called KeYmaera. Once we adapt the model to a hybrid system representation we proceed to use KeYmaera to verify that the proposed model is formally correct.
\end{abstract}


\section{Introduction}

This work will focus on the representation of a vehicular traffic network model using hybrid systems, where the model to represent is given in ordinary differential equations and works on a macroscopic scale, that is, the variable to study is the flow of traffic and not each vehicle individually. Furthermore, this model describes vehicular traffic through three events and in two of these the interaction of traffic lights appears. \\

Now what are hybrid systems? Hybrid systems are systems that describe the interaction of the continuous with the discrete that appears more and more in the devices or in the events where decisions have to be made, for example an airplane, this can be described from a physical theory, but this theory fails to accurately describe the moment in which the aircraft must make decisions, such as when a possible collision, this is where computerized systems determine what action the plane should take. The example shows a device in this case an airplane that can be described from the continuous (physical) or from the discrete (computerized), but it would be better to be able to describe it taking into account both parts, that is why begins to study these hybrid systems from different scopes. \\

For this work we will take the approximation of  Platzer \cite{Andre}, where the hybrid systems are represented from special graphs called hybrid automaton, then with the help of the differential dynamics logic. These automaton can be transformed into hybrid programs, in order to finally be able to formally verify each of the possible states achievable by the chosen model. This allows a more exhaustive verification than if the model were to be verified through a numerical method.

Section 3 is dedicated to review some fundamental ideas on hybrid systems along with the presentation of dynamic differential logic and the use of the KeYmaera tool. . In this section we also discuss a model proposed in \cite{Argen} which will become the system to be tested with the proof theoy of dynamic differential logic .
Section 4 is directed to the construction of a hybrid system based on the ordinary equations transport model. With it, we discuss the formal verification of this model through  of KeYmaera.

\section{Preliminaries}
\subsection{Hybrid Automaton}
We will see a formal definition of what the automaton mentioned above are and the relationship they have with hybrid programs.

\begin{definition}	
	A hybrid automaton A consists in:
	\begin{itemize}
		\item A state space $\mathbb{R}^n$.
		
		\item A finite directed graph with vertices Q (as modes) y edges E (as control switches).
		
		\item Flow conditions $\textit{flow}_{q} \subseteq \mathbb{R}^n \times \mathbb{R}^n$ that determine the relationship of the state $x \in \mathbb{R}^n$ and its time derivative  $x' \in \mathbb{R}^n$ during continuos evolution in mode q $\in$ Q. 
		
		\item Invariant condition $\textit{inv}_q \subseteq \mathbb{R}^n$ or evolution domain restrictions that have to be true while in mode q $\in$ Q.
		
		\item Jump relation $\textit{jump}_e \subseteq \mathbb{R}^n \times \mathbb{R}^n $ that determine the new value of the state x $\in \mathbb{R}^n$ depending on its old value when following edge e $\in$ E. 
	\end{itemize}
\end{definition}
In order to understand the states of the automaton, it is necessary to make a semantic transition, and thus interpret them properly.

\begin{definition}
	\textbf{[Transition semantics of hybrid automata]}. The transition system of a hybrid automaton A is a transition relation $\curvearrowright$ defined as follow:
	
	\begin{itemize}
		\item The state space is defined as S := \{(q,x) $\in$ Q $\times$    $\mathbb{R}^n$ : \mbox{x $\in$ $\textit{inv}_q$}\}.
		
		\item The transition relation $\curvearrowright \hspace{0.1cm} \subseteq$ S $\times$ S is defined as the union $\bigcup_{e\in E} \curvearrowright^e \hspace{0.1cm} \cup \hspace{0.1cm} \bigcup_{q \in Q} \curvearrowright^q$ where:
		\begin{enumerate}
			\item (q,x) $\curvearrowright^e$ ($\hat{q},\hat{x}$) iff e $\in$ E is an edge from q $\in$ Q a $\hat{q} \in $ Q in the hybrid automaton and (x,$\hat{x}$) $\in \textit{jump}_e$(discrete transition).
			
			\item  (q,x) $\curvearrowright^q$ ($\hat{q},\hat{x}$) iff q $\in$ Q and there is a function f:[0,r]$\longrightarrow \mathbb{R}^n$ that has a time derivative $f'$:(0,r)$\longrightarrow \mathbb{R}^n$ such that f(0) = x, f(r) = $\hat{x}$ and that respects (f($\zeta$),$f'(\zeta$))$\hspace{0.1cm} \in \hspace{0.1cm}  \textit{flow}_q$ at each $\zeta \in $ (0,r). Further, f($\zeta$)$\hspace{0.1cm} \in \hspace{0.1cm} \textit{inv}_q$ for each $\zeta \in$[0,r](continuous transition). 
		\end{enumerate} 
	\end{itemize}
	State $\sigma \in$ S is reachable from state $\sigma_0 \in $ S, denoted by $\sigma_0 \curvearrowright^* \sigma$, iff, for some $\textit{n} \in \mathbb{N}$, there is a sequence of states $\sigma_1, \sigma_2,.....,\sigma_n \in $ S such that $\sigma_{i-1} \curvearrowright \sigma_i$ for \mbox {1$\leq$ i $\leq$ n.}
\end{definition}

One of the advantages of using hybrid automaton is that they can be faithfully represented as a hybrid program, that is, a computer program that expresses the same relationship that defines the automated system and thus use software in order to verify the correctness of the implementation made. However, for the correctness verification, it is necessary to define a transition of the automaton program and validate that the continuous-discrete relation of the automaton is not lost in said transition.

The following statement establishes the equivalence of the automaton-program relationship.

\begin{proposition} \textbf{[Hybrid automata embedding]}.
	There is an effective mapping $\iota$ from hybrid automata to hybrid programs such that the following diagram commutes:
	
	\begin{figure}[H]
		\centering
		\includegraphics[width=0.7\linewidth]{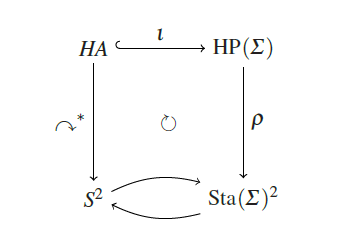}
		\caption{Source:\cite{Andre} }
		\label{fig:diaconm}
	\end{figure}
	
	That is, the transition semantics $ \rho(\iota(A)) $ of the hybrid program $ \ iota (A) $ corresponding to the hybrid automaton A, this transition is identical to the reachability relation $ \curvearrowright^*$ corresponding to the  automaton A when states of the hybrid program in $Sta(\Sigma)^2$ are identified with states of the hybrid automaton in S by canonical bijection.
\end{proposition}
The proof of this proposition can be found in \cite{Andre} pages 372-373.

The proposition states that the hybrid program can be rewritten and interpreted from the logic d$L$.

\begin{corollary}
	There is an effective mapping from safety properties of a hybrid automata to d$L$ formulas such that the hybrid automaton A, starting at mode $q_0$, stays safely in the region of F $\in Fml_{FOL} (\Sigma)$ if and only if the corresponding d$L$ formula is valid.
\end{corollary}

This corollary means that as long as the d$L$ formulas are valid, there will be a safe path where the properties of the automaton and its region of evolution will be maintained in passing to a hybrid program. \\

As it is necessary to express the programs in logic d$L$, then we will see its definition below. \\

\subsection{Differential Dynamic Logic}
Now we will introduce the differential dynamic logic (d$L$) in which operational models of hybrid systems are introduced as first-class citizens, that is, syntactically manipulable through of a convenient grammar. This allows the transitions of the behavior of the hybrid system to be expressed as formulas as long as they are correct. \\

As a basis, d$L$ includes real arithmetic to describe concepts such as safe regions of the state space, it also has support for real-value quantifiers, to quantify possible values of parameters or variables changing over time. For the behavioral transitions of systems, d$L$ makes use of modal operators, such as [$\alpha$] or $\langle \alpha \rangle$ that refer to the states achievable by the $\alpha$ program. \cite{Andre} \\

This logic has its own syntax, semantics, and sequence calculus, as we will show below. \\

Let $ V $ be the set of all logical variables, and $ \Sigma $ the set of all symbols or signature. To establish the syntax, it is necessary to define terms, first-order formulas, dynamic differential formulas, and hybrid programs.

\begin{definition} \textbf{[Terms]}.
	Trm($\Sigma,V$) is the set of all terms, which is the smallest set such that:
	\begin{itemize}
		\item If x $\in V$, then x $\in$ Trm($\Sigma$,$V$).
		
		\item If f $\in \Sigma$ is a function symbol of arity n $\geq$ 0, and for \mbox {1 $\leq$ i $\leq$ n}, $\theta_i \in$ Trm($\Sigma$,$V$), then f($\theta_1,.....,\theta_n$) $\in$ Trm($\Sigma$,$V$). The case $n=0$ is permitted.  
	\end{itemize}
\end{definition}

Terms are well-formed arguments that run in symbol or predicate functions. For example, logical variables are well-formed terms.

Every formula has a sense of truth or falsehood. In a given context these formulas are called well-formed formulas. A formula consists of all the words that can be recursively formed by combining symbols of the signature with appropriate logical operators.

\begin{definition} \textbf{[First-order fomulas]}.
	The set $Fml_{FOL}(\Sigma, V)$ of formulas of first-order logic is the smallest set with:
	
	\begin{itemize}
		\item If p $\in \Sigma$ is a predicate symbol of arity n $\geq$ 0 and $\theta_i \in$ Trm($\Sigma$,$V$) for \mbox {1 $\leq$ i $\leq$ n}, then p($\theta_1,...,\theta_n$) $\in Fml_{FOL}(\Sigma,V)$.
		
		\item If $\phi , \psi \in Fml_{FOL}(\Sigma,V)$, then  $\neg \phi , (\phi \wedge \psi), (\phi \lor \psi), (\phi \rightarrow \psi), \in Fml_{FOL}(\Sigma,V)$.
		
		\item If $\phi \in Fml_{FOL}(\Sigma,V)$ and  x $\in$ V, then $(\forall x \phi), (\exists x \phi) \in Fml_{FOL}(\Sigma,V)$. 
	\end{itemize}
\end{definition} 

As there is a rewriting theorem between automaton and programs, it is necessary to clearly define what we understand as a program.

\begin{definition}
	\textbf{[Hybrid programs]}. The set HP($\Sigma$,$V$) of hybrid programs, with the typical elements $\alpha,\beta$, is defined inductively as the smallest set such that:
	
	\begin{enumerate}
		\item If $x_i \in \Sigma$ is a state varible and $\theta_i \in$ Trm($\Sigma$,$V$) for \mbox {1 $\leq$ i $\leq$ n}, then the discrete jump set ($x_1:= \theta_1,...,x_n:= \theta_n )\in HP(\Sigma,V)$ is a hybrid program. We assume that the $x_1,...,x_n$ are pairwise different state variables.
		
		\item If $x_i \in \Sigma$ is a state variable and $\theta_i \in$ Trm($\Sigma$,$V$) for \mbox {1 $\leq$ i $\leq$ n}, then $x'_i = \theta_i$ is a differential equation in which $x'_i$ represents the time derivative of variable $x_i$. If $\chi
		$ is a first-order formula, then ($x'_1:= \theta_1,...,x'_n:= \theta_n \& \chi )\in HP(\Sigma,V)$. We assume that $x_1,...,x_n$ are pairwise different stat variables.
		
		\item If $\mu$ is a first-order formula, then (?$\mu$) $\in HP(\Sigma,V)$. 
		
		\item If $\alpha, \beta \in HP(\Sigma,V)$, then $(\alpha \cup \beta) \in HP(\Sigma,V)$.
		
		\item If $\alpha, \beta \in HP(\Sigma,V)$, then $(\alpha;\beta) \in HP(\Sigma,V)$.
		
		\item If $\alpha \in HP(\Sigma,V)$, then $(\alpha^*) \in HP(\Sigma,V)$.
		
	\end{enumerate}
\end{definition}



Now we will define the formulas for d$L$.

\begin{definition}
	\textbf{[d$L$ Formulas]}. The set Fml($\Sigma,V$) of formulas of d$L$, with typical elements $\alpha, \beta$, is the smallest set such that:
	
	\begin{enumerate}
		\item If p is a predicate symbol of arity n $\geq$ 0 and  $\theta_i \in$ Trm($\Sigma$,V) for \mbox {1 $\leq$ i $\leq$ n}, then p($\theta_1,..,\theta_n) \in Fml(\Sigma,V) $.
		
		\item   If $\phi , \psi \in Fml(\Sigma,V)$, then  $\neg \phi , (\phi \wedge \psi), (\phi \lor \psi), (\phi \rightarrow \psi), \in Fml(\Sigma,V)$.
		
		\item If $\phi \in Fml(\Sigma,V)$ y x $\in$ $V$, then $(\forall x \phi), (\exists x \phi) \in Fml(\Sigma,V)$.
		
		\item If $\phi \in Fml(\Sigma,V)$ y $\alpha \in HP(\Sigma,V)$, then $[\alpha]\phi, \langle \alpha \rangle \phi \in Fml(\Sigma,V)$.
	\end{enumerate}
\end{definition}

Defined the syntax of d$L$, next we will define the semantics. Therefore, we will define the valuation of the terms and formulas and the semantic transition of the programs.

\begin{definition}
	\textbf{[Valuation of terms]}. The valuation of the terms with respect to interpretation I, assignment $\eta$, and state v is defined by:
	\begin{enumerate}
		\item $\textit{val}_{I,\eta}(v,x) = \eta(x)$ if x $\in$ V is a logic variable.
		
		\item  $\textit{val}_{I,\eta}(v,a) = v(a)$ if a $\in \Sigma$  is a state variable.
		
		\item $\textit{val}{I,\eta}(v,f(\theta_1,...,\theta_n)) = I(f)(\textit{val}{I,\eta}(v,\theta_1),....,\textit{val}_{I,\eta}(v,\theta_n))$ when f $\in \Sigma$ is a rigid function symbol of arity n $\geq$ 0.
		
	\end{enumerate}
\end{definition}

\begin{definition}
	\textbf{[Valuation of d$L$ formulas]}. The valuation $\textit{val}_{I,\eta}(v,.)$ of formulas with respect of the interpretation I, assigment $\eta$, and state v is defined by:
	
	\begin{enumerate}
		\item $\textit{val}{I,\eta}(v,p(\theta_1,...,\theta_n)) = I(p)(\textit{val}{I,\eta}(v,\theta_1),....,\textit{val}_{I,\eta}(v,\theta_n))$.
		
		\item $\textit{val}{I,\eta}(v,\phi \wedge \psi) = \textit{true}$ iff $\textit{val}{I,\eta}(v,\phi) = \textit{true}$ and $\textit{val}_{I,\eta}(v,\psi) = \textit{true}$.
		
		\item $\textit{val}{I,\eta}(v,\phi \lor \psi) = \textit{true}$ Iff $\textit{val}{I,\eta}(v,\phi) = \textit{true}$ or $\textit{val}_{I,\eta}(v,\psi) = \textit{true}$.
		
		\item $\textit{val}{I,\eta}(v,¬\phi) = \textit{true}$ iff $\textit{val}{I,\eta}(v,\phi) \neq \textit{true}$.
		
		\item $\textit{val}{I,\eta}(v,\phi \rightarrow \psi) = \textit{true}$ iff $\textit{val}{I,\eta}(v,\phi) \neq \textit{true}$ or$\textit{val}_{I,\eta}(v,\psi) = \textit{true}$.
		
		\item $\textit{val}{I,\eta}(v,\forall x \phi) = \textit{true}$ iff $\textit{val}{I,\eta[x\rightarrow d]}(v,\phi) = \textit{true}$  for all d $\in \mathbb{R}$.
		
		\item $\textit{val}{I,\eta}(v,\exists x \phi) = \textit{true}$ iff $\textit{val}{I,\eta[x \rightarrow d]}(v,\phi) = \textit{true}$ for some d $\in \mathbb{R}$.
		
		\item $\textit{val}{I,\eta}(v,[\alpha]\phi) = \textit{true}$ iff $\textit{val}{I,\eta}(\omega,\phi) = \textit{true}$ for all states $\omega$ for which the transition relation satisfies $(v,\omega) \in \rho_{I,\eta}(\alpha)$.
		
		\item $\textit{val}{I,\eta}(v,\langle \alpha \rangle \phi) = \textit{true}$ iff $\textit{val}{I,\eta}(\omega,\phi) = \textit{true}$ for some state $\omega$ for which the transition relation satisfies $(v,\omega) \in \rho_{I,\eta}(\alpha)$.

	\end{enumerate}
\end{definition}
Following the usual notation, we can also write I, $\eta$, v $\models \phi$ if and only if $\textit{val}_{I, \eta}=$ true. Then we say that $\phi$ is satisfiable at I, $\eta$, v or is true at I, $\eta$, v. We also say that I, $\eta$, v is a model of $\phi$. If $\phi$ is satisfied for at least one of I, $\eta$, v then $\phi$ is called satisfied. Occasionally we only write $\vDash \phi$ if and only if I, $\eta$, v $\models \phi$ for all I, $\eta$, v. Then the formula $\phi$ is called valid, that is, true for all assignments I, $\eta$, v.
\begin{definition}
	\textbf{[Transition semantics of hybrid programs ]}. The valuation of a hybrid program $\alpha$, denoted by $\rho_{I,\eta}(\alpha)$, it is a transition relation on states. It specifies which state $\omega$ is reachable from a state v by operations of hybrid program $\alpha$ and is defined as follows:
	
	\begin{enumerate}
		\item $(v,\omega) \in \rho_{I,\eta}(x_1:=\theta_1,.....,x_n:=\theta_n)$ if and only if the state $\omega$ equals the state obtained by semantic modification of the state v as v[$x_1 \rightarrow \textit{val}{I,\eta}(v,\theta_1),.....,x_n \rightarrow \textit{val}{I,\eta}(v,\theta_n)$]. Particularly, the values of other variables z $\notin \{x_1,...,x_n\} $ remain constant, which means, $\textit{val}{I,\eta}(\omega,z) = \textit{val}{I,\eta}(v,z)$, and the $x_i$ receive their new values simultaneously, which means, $\textit{val}{I,\eta}(\omega,x_i) = \textit{val}{I,\eta}(v,\theta_i)$.
		
		\item $(v,\omega) \in \rho_{I,\eta}(x'_1:=\theta_1,.....,x'_n:=\theta_n \hspace{0.1cm}  \& \hspace{0.1cm} \chi)$ if and only if there is a flow f of some duration r $\geq$ 0 from state v to state $\omega$ along $x'_1 = \theta_1,....,x'_n = \theta_n \hspace{0.1cm} \& \hspace{0.1cm} \chi$, which means, a function $f:[0,r] \rightarrow Sta(\Sigma)$ such that:
		
		\begin{itemize}
			\item $f(0) = v, f(r) = \omega$.
			
			\item f respects the differential equations: For each variable $x_i$, the valuation $\textit{val}{I,\eta}(f(\zeta),x_i) = f(\zeta)(x_i)$ of $x_i$ in the state $f(\zeta)$ is continuous in $\zeta$ on [0,r] and has derivative of value $\textit{val}{I,\eta}(f(\zeta),\theta_i)$ at each time $\zeta \in$ (0,r).
			
			\item The value of other variables z $\notin \{x_1,....,x_n\}$ remain constant, that is, we have $\textit{val}{I,\eta}(f(\zeta),z) = \textit{val}{I,\eta}(v,z)$ for all $\zeta \in$ [0,r].
			
			\item f respects the invariant: $\textit{val}_{I,\eta}(f(\zeta),\chi) = \textit{true}$ for each $\zeta \in$ [0,r].
		\end{itemize} 
		
		\item $\rho_{I,\eta}(?\mu) = \{(v,v) : \textit{val}_{I,\eta}(v,\mu) = \textit{true}\}$.
		
		\item $\rho_{I,\eta}(\alpha \cup \beta) = \rho_{I,\eta}(\alpha) \cup \rho_{I,\eta}(\beta)$.
		
		\item $\rho_{I,\eta}(\alpha ; \beta) = \{(v,\omega) : (v,\mu) \in \rho_{I,\eta}(\alpha), (\mu, \omega) \in \rho_{I,\eta}(\beta) \hspace{0.2cm} \text{for all} \hspace{0.2cm}\zeta \in [0,r]\}$.
		
		\item $(v,\omega) \in \rho_{I,\eta}(\alpha^*)$, iff there is an n $\in \mathbb{N}$ and states $v = v_0, v_1,...v_{n-1},v_n = \omega$ such that $(v_i,v_{i+1}) \in \rho_{I,\eta}(\alpha)$ for all \mbox{0 $\leq$ i $\leq$ n.}  
	\end{enumerate}
\end{definition}

Since we are working with differential equations, we have to make sure that their solutions are unique. 

\begin{lemma}
	\textbf{Uniqueness}. The differential equations of d $L$ have a unique solution, that is, for each system of differential equations, and each state v, and each duration r $\geq$ 0, there is at most one flow $f:[0, r] \rightarrow Sta(\Sigma)$ that satisfies Case 2 of Definition 9.
\end{lemma}

Now we will discuss the notion of substitution which will be fundamental in the calculation of sequences.

\begin{definition}
	\textbf{[Admissible substitution]}. An application of a substitution $\sigma$  is permissible if there is no x variable that replaces $\sigma$ with $\sigma(x)$ occurring within a quantifier or a modality binding x to a variable of the replacement $\sigma(x)$. A modality binds a state variable x if and only if it contains a discrete jump set assigning to x or a differential equation containing $x'$.
\end{definition}

\begin{lemma}
	\textbf{[Substitution Lemma]}. Let $\sigma$ an admissible substitution for the formula $\phi$, and let $\sigma$ replace only logical variables, then:
	$$\text{para cada} \hspace{0.2cm} I,\eta, v: \textit{val}{I,\eta}(v,\sigma(\phi)) = \textit{val}{I,\sigma^*(\eta)}(v,\phi)$$
	where the semantic modification $\sigma^(\eta)$ of assigment $\eta$ is adjoint to $\sigma$, which means, $\sigma^(\eta)$ is identical to $\eta$, except that $\sigma^*(\eta)(x) = \textit{val}_{I,\eta}(v,\sigma(x))$ for all logical variables x $\in$ V.
\end{lemma}
The proof of this lemma is in \cite{Andre} pages 70-74.



\begin{lemma}
	\textbf{[Substitutions preserve validity]}. If $\models \phi$, that is, $\phi$ is valid, then $\models \sigma(\phi)$ for any substitution $\sigma$ that is admissible for $\phi$.
\end{lemma}
The proof of this lemma is in \cite{Andre} page 76.\\

When calculating sequences we must think about how to eliminate the quantifiers and be able to leave the variables linked to those quantifiers free, that is why the following definition is introduced:

\begin{definition}
	\textbf{[Quantifiers elimination]}. A first-order theory admits elimination of quantifiers if, in each $\phi$ formula, an equivalent QE formula ($\phi$) free of quantifiers can be effectively associated, that is that is, $\phi \leftrightarrow QE (\phi)$ is valid. Also this formula should not have more formulas with free variables or symbolic functions.
	
\end{definition}

Now we will introduce the rules of sequences calculation.

\begin{enumerate}
	
	\item Basic Rules

	\begin{multicols}{2}
		\begin{prooftree}
			\AxiomC{$\phi \vdash$}
			\LeftLabel{$\lnot r$}
			\UnaryInfC{$\vdash \lnot  \phi$}
		\end{prooftree}
		
		\columnbreak
		
		\begin{prooftree}
			\AxiomC{$\vdash \phi , \psi$}
			\LeftLabel{$\wedge r$}
			\UnaryInfC{$\vdash \phi \wedge \psi $}
		\end{prooftree}  
	\end{multicols}
	
	\begin{multicols}{2}
		\begin{prooftree}
			\AxiomC{$\vdash \phi$}
			\LeftLabel{$\lnot l$}
			\UnaryInfC{$\lnot \phi \vdash$}
		\end{prooftree} 
		\columnbreak
		\begin{prooftree}
			\AxiomC{$\phi \vdash$}
			\AxiomC{$\psi \vdash$}
			\LeftLabel{$\wedge l$}
			\BinaryInfC{$\phi \wedge \psi \vdash$}
		\end{prooftree} 
	\end{multicols}
	
	\begin{multicols}{2}
		\begin{prooftree}
			\AxiomC{$\vdash \phi$}
			\AxiomC{$\vdash \psi$}
			\LeftLabel{$\lor r$}
			\BinaryInfC{$\vdash \phi \lor \psi$}
		\end{prooftree} 
		
		\columnbreak
		
		\begin{prooftree}
			\AxiomC{$\phi \vdash \psi$}
			\LeftLabel{$\rightarrow r$}
			\UnaryInfC{$\vdash \phi \rightarrow \psi $}
		\end{prooftree} 
	\end{multicols}
	
	\begin{multicols}{2}
		
		\begin{prooftree}
			\AxiomC{$\phi , \psi \vdash$}
			\LeftLabel{$\lor l$}
			\UnaryInfC{$\phi \lor \psi\vdash$}
		\end{prooftree}
		\columnbreak
		\begin{prooftree}
			\AxiomC{$\phi \vdash$}
			\AxiomC{$\psi \vdash$}
			\LeftLabel{$\rightarrow l$}
			\BinaryInfC{$\phi \rightarrow \psi \vdash$}
		\end{prooftree}
	\end{multicols}
	
	\item Axiom and Cut
	
	\begin{multicols}{2}
		\begin{prooftree}
			\AxiomC{}
			\LeftLabel{ax}
			\UnaryInfC{$\phi \vdash \phi$}
		\end{prooftree}  
		
		\columnbreak
		\begin{prooftree}
			\AxiomC{$\vdash \phi$}
			\AxiomC{$\phi \vdash$}
			\LeftLabel{cut}
			\BinaryInfC{$\vdash$}
		\end{prooftree} 
		
	\end{multicols}
	\item Dynamic Rules
	
	\begin{multicols}{3}
		\begin{prooftree}
			\AxiomC{$\langle \alpha \rangle \langle \beta \rangle \phi$}
			\LeftLabel{$\langle ; \rangle$}
			\UnaryInfC{$\langle \alpha ; \beta \rangle \phi$}
		\end{prooftree} 
		
		\columnbreak
		
		\begin{prooftree}
			\AxiomC{$\phi \lor \langle \alpha \rangle \langle \alpha^*\rangle \phi$}
			\LeftLabel{$\langle ^{*n} \rangle $}
			\UnaryInfC{$\langle \alpha^*\rangle \phi$}
		\end{prooftree} 
		
		\columnbreak
		
		\begin{prooftree}
			\AxiomC{$\langle\alpha\rangle \phi \lor \langle\beta\rangle \phi$}
			\LeftLabel{$\langle \cup \rangle$}
			\UnaryInfC{$\langle\alpha \cup \beta\rangle \phi$}
		\end{prooftree}
	\end{multicols}
	
	\begin{multicols}{3}
		\begin{prooftree}
			\AxiomC{$[ \alpha ] [ \beta ] \phi$}
			\LeftLabel{$[;]$}
			\UnaryInfC{$[ \alpha ; \beta ]\phi$}
		\end{prooftree} 
		
		\columnbreak
		
		\begin{prooftree}
			\AxiomC{$\phi \wedge [ \alpha ] [ \alpha^*] \phi$}
			\LeftLabel{$[^{*n}] $}
			\UnaryInfC{$[ \alpha^*]\phi$}
		\end{prooftree} 
		
		\columnbreak
		
		\begin{prooftree}
			\AxiomC{$[\alpha] \phi \wedge [\beta] \phi$}
			\LeftLabel{$[ \cup ]$}
			\UnaryInfC{$[\alpha \cup \beta] \phi$}
		\end{prooftree}
	\end{multicols}
	
	\begin{multicols}{2}
		\begin{prooftree}
			\AxiomC{$\chi \wedge\psi$}
			\LeftLabel{$\langle  ? \rangle$}
			\UnaryInfC{$\langle?\chi\rangle \psi$}
		\end{prooftree}
		\columnbreak
		\begin{prooftree}
			\AxiomC{$\chi \rightarrow \psi$}
			\LeftLabel{$[  ? ]$}
			\UnaryInfC{$[?\chi] \psi$}
		\end{prooftree}
	\end{multicols}
	
	\item Substitution
	
	\begin{multicols}{2}
		\begin{prooftree}
			\AxiomC{$\phi_{x_1}^{\theta_1}...\phi_{x_n}^{\theta_n}$}
			\LeftLabel{$\langle:=\rangle$}
			\UnaryInfC{$\langle x_1:=\theta_1,...,x_n:=\theta_n\rangle \phi$}
		\end{prooftree} 
		\columnbreak
		\begin{prooftree}
			\AxiomC{$\langle x_1:=\theta_1,...,x_n:=\theta_n\rangle \phi$}
			\LeftLabel{$[:=]$}
			\UnaryInfC{$[ x_1:=\theta_1,...,x_n:=\theta_n]\phi$}
		\end{prooftree} 
		
	\end{multicols}
	\item Dynamic Rules for Differential Equations
	
	\begin{prooftree}
		\AxiomC{$\exists t \geq 0((\forall 0\leq \hat{t} \leq t \langle \xi_{\hat{t}}\rangle \chi) \wedge \langle \xi_t \rangle \phi)$}
		\LeftLabel{$\langle ' \rangle$}
		\UnaryInfC{$\langle x'_1=\theta_1,...,x'_n=\theta_n \& \chi \rangle \phi$}
	\end{prooftree} 
	
	\begin{prooftree}
		\AxiomC{$\forall t \geq 0((\forall 0\leq \hat{t} \leq t \langle \xi_{\hat{t}}\rangle \chi) \rightarrow \langle \xi_t \rangle \phi)$}
		\LeftLabel{$[ ' ]$}
		\UnaryInfC{$[ x'_1=\theta_1,...,x'_n=\theta_n \& \chi ] \phi$}
	\end{prooftree} 
	
	
	\item Quantifiers Elimination
	
	\begin{multicols}{2}
		\begin{prooftree}
			\AxiomC{$\vdash \phi(s(X_1,...,X_n))$}
			\LeftLabel{$\forall r$}
			\UnaryInfC{$\vdash \forall x \phi(x)$}
		\end{prooftree} 
		\columnbreak
		\begin{prooftree}
			\AxiomC{$\vdash \phi(X)$}
			\LeftLabel{$\exists r$}
			\UnaryInfC{$\vdash \exists x \phi(x)$}
		\end{prooftree} 
	\end{multicols}
	
	\begin{multicols}{2}
		\begin{prooftree}
			\AxiomC{$ \phi(X) \vdash$}
			\LeftLabel{$\forall l$}
			\UnaryInfC{$\forall x \phi(x) \vdash$}
		\end{prooftree}
		\columnbreak
		\begin{prooftree}
			\AxiomC{$\phi(s(X_1,...,X_n)) \vdash$}
			\LeftLabel{$\exists l$}
			\UnaryInfC{$\exists x \phi(x) \vdash$}
		\end{prooftree}
	\end{multicols}
	
	\item Quantifiers Introduction
	
	\begin{prooftree}
		\AxiomC{$\vdash QE(\forall X(\Phi (X) \vdash \Psi(X)))$}
		\LeftLabel{$i \forall$}
		\UnaryInfC{$\Phi(s(X_1,...,X_n)) \vdash \Psi(s(X_1,....,X_n))$}
	\end{prooftree}  
	
	\begin{prooftree}
		\AxiomC{$\vdash QE(\exists X \bigwedge_i (\Phi_i \vdash \Psi_i))$}
		\LeftLabel{$i \exists$}
		\UnaryInfC{$\Phi_1\vdash \Psi_1 .... \Phi_n \vdash \Psi_n$}
	\end{prooftree} 
	
	\item Global Rules
	
	\begin{multicols}{2}
		\begin{prooftree}
			\AxiomC{$\vdash \forall^{\alpha}(\phi \rightarrow \psi)$}
			\LeftLabel{[] gen}
			\UnaryInfC{$[\alpha]\phi \vdash [\alpha] \psi$}
		\end{prooftree} 
		\columnbreak
		\begin{prooftree}
			\AxiomC{$\vdash \forall^{\alpha}(\phi \rightarrow \psi)$}
			\LeftLabel{$\langle\rangle$ gen}
			\UnaryInfC{$\langle\alpha\rangle\phi \vdash \langle\alpha\rangle \psi$}
		\end{prooftree} 
	\end{multicols}
	
	\begin{multicols}{2}
		\begin{prooftree}
			\AxiomC{$\vdash \forall^{\alpha}(\phi \rightarrow [\alpha]\phi)$}
			\LeftLabel{ind}
			\UnaryInfC{$\phi \vdash [\alpha^*]\phi $}
		\end{prooftree}
		\columnbreak
		\begin{prooftree}
			\AxiomC{$\vdash \forall^{\alpha} \forall v>0 (\varphi(v) \rightarrow \langle\alpha\rangle \varphi(v-1))$}
			\LeftLabel{con}
			\UnaryInfC{$\exists v \varphi(v) \vdash \langle\alpha\rangle \exists v \leq 0 \varphi(v)$}
		\end{prooftree}
	\end{multicols}

\end{enumerate}

The $\neg r$ -cut rules are the standard propositional rules, they decompose the propositional structure of the formulas. The rules $\neg r $ and $\neg l$ use simple dualities caused by semantic implications in the sequences. The $\vee r$ rule uses the notion that formulas are disjunctively combined in sequents, the $\wedge l$ rule uses the same notion but with the antecedents, that is, these are joined by conjunctions. The rules $\vee l$ and $\wedge r$ divide the proof into two cases, because the conjunction in sequents can be tested separately, as well as the disjunction in the antecedents. The $\rightarrow r$ rule uses the fact that the implication has the same meaning as the sequence arrow $\vdash$. The rule $\rightarrow l$ is divided in two ways because we do not know if the implication in the antecedent is fulfilled.

The axiom rule (ax) closes the result and the cut rule assumes the appearance of any additional formula to the left or to the right, that is, in the antecedents or in the sequents.

The rule $\exists l$ wants to test $\exists x \phi(x) $ in the sequence, for this it introduces a free variable X for the variable x that was being quantized by an exists. Its dual is the rule $ \forall l $, which assumes $\forall x \phi(x) $ in the antecedent, it introduces a new free logical variable X for the variable x that was being quantized by a para everything.

The rule $ \forall r $, where we want to test $ \forall x \phi(x) $ in the sequence, for this we introduce a new symbolic function s for the quantized variable x and replace this variable with a term  s ($ X_1, X_2, ...., X_n $) where $ X_1, X_2, ...., X_n $ are free logical variables of the original formula $ \forall x \phi(x) $. The rule $ \exists r $ is very similar because we assume $ \exists x \phi(x) $ in the antecedent, but we only know that that x exists, but not the value it has, so a symbolic function is introduced and replace x by s ($ X_1, ...., X_n $), where $ X_1, ...., X_n $ are free logical variables.

With the $ i \forall $ rule we can reintroduce a universal quantifier for a term of the form s ($ X_1, ..., X_n $), which corresponds to a variable that was previously being quantized in the antecedent. The dual rule $ i \exists $ can reintroduce an existential quantifier for a free logical variable that was previously quantized in the sequent or universally quantized in the antecedent.

The rules of dynamic modality transform the hybrid programs into logical formulas with simpler structures through symbolic decomposition. First there are the non-deterministic choice rules like ($ \langle \cup \rangle, [\cup] $). For the rule [$ \cup $] if all $ \alpha $ transitions lead to states satisfying $ \phi $ and all $ \beta $ transitions lead to states satisfying $ \phi $, then all program transitions $ \alpha \cup \beta $ also lead to states that satisfy $ \phi $. Dually for the rule $ \langle \cup \rangle $, if there is a transition  $ \alpha $ to state $ \phi $ or if there is a transition  $ \beta $ to state $ \phi $, then in any case there is a transition from $ \alpha \cup \beta $ to $ \phi $.
The sequential composition rules ($ \langle; \rangle, [;] $), for the rule [;] if all transitions $ \alpha $ and all transitions $ \beta $ lead to states that satisfy a $ \phi $ then the sequential composition of them will also make it. The rule $ \langle; \rangle $ uses the fact that if there is a transition on $ \alpha $ and there is a transition on $ \beta $ leading to $ \phi $, then there is a sequential composition that also reaches $ \phi $.

The rules ($ \langle ^{*n} \rangle, [^{*n}] $), are the usual iteration rules, which unwrap loops. The rules ($ \langle? \rangle, [?] $) Are questions that are tested showing that this question can be solved, that is, if $? \chi $ can only make the transition when the condition  $ \chi $ stays true. The substitution rules ($ \langle: = \rangle, [: =] $), are there to make discrete replacements when both values are true.

The last rule block [] gen, $ \langle \rangle $ gen, ind, with are global rules. These depend on the truth of the premises in all their states attainable by the hybrid program  $ \alpha $, for which it is ensured that the universal lock $ \forall^{\alpha} $ with respect to all state variables bundles of the respective hybrid program  $ \alpha $. This universal lock over-approximates all possible $ \alpha $ changes, since it encompasses all bound variables. This universal lock is necessary for the validity of the presence of contexts $ \Gamma, \Delta $ or of free variables.

\begin{definition}
	\textbf{[Provability]}. A derivation is a finite, acylic, sequent-labeled graph such that, for each node, the labels of its children must be instances of one of the calculation rules and the labels of the parents of these children must be the conclusion of the instance of that rule. A formula $ \psi $ is probable from a set $ \Phi $ of formulas, denoted by $ \Phi \vdash_{dL} \psi $ if and only if there exists a finite subset $ \Phi_0 \subseteq \Phi $ for each sequent $ \Phi_0 \vdash \psi $ is derivable.
\end{definition}



\begin{lemma}
	\textbf{[Coincidence lemma]}. If the interpretations and assigments and states, I,$\eta$,v respectively and J,$\epsilon , \omega$ agree on all symbols that occur freely in the formula $\phi$ then $\textit{val}{I,\eta}(v,\phi)=\textit{val}{J,\epsilon}(\omega,\phi)$
\end{lemma}

Now we must ask if d$L$ logic is valid and complete.

\begin{theorem}
	\textbf{[Soundness]}. Differential dynamic logic is sound or valid, which means, all instances of the states are valid.
\end{theorem}
The proof of this theorem is in \cite{Andre} pages 98-101. 

\begin{theorem}
	\textbf{[Incompleteness of d$L$]}. Both the discrete part and the continuous part of d $L$ are not effectively axiomatizable, that is, they do not have a solid and complete calculation, since the natural numbers can be defined as discrete or as continuous.
\end{theorem}
The proof of this theorem is in \cite{Andre} pages 102-103.

\begin{theorem}
	\textbf{[Relative completeness of d$L$]}. The calculation of d$L$ is relatively complete to FOD (First Order Differential Equations of Logic), that is, every valid d$L$ formula can be derived from FOD tautologies .
	
\end{theorem}

The proof of this theorem is in \cite{Andre} page 104.




\subsection{KeYmaeraX}
KeYmaeraX is an interactive theorem tester. Its input is a formula of dynamic differential logic, combining both the description of the system and the properties under consideration. To test this formula, it is divided into several sub-arguments according to the $dL$ sequence rules. The Boolean structure of the input formula is correctly transformed into a test tree. The programs are carried out by symbolic execution, that is, for each program that is built there is a test rule that calculates its effect. For example, the assignment x $:= \theta $ can be executed by replacing each occurrence of x with the new value $ \theta $. Likewise, decisions in program flow can be explored separately, that is, $ [\alpha \cup \beta] \theta $ is true if and only if $ [\alpha] \theta $ and $ [\beta ] \theta $ are true, because they are possible paths, so the system $ \alpha \bigcup \beta $ can only be safe if all $ \alpha $ executions and all $ \beta $ executions are safe . KeYmaeraX uses inductive invariants for loops. An inductive invariant to test $ \theta \rightarrow [\alpha^*] \psi $ is a formula $ J $ that satisfies the current state ($ \theta \rightarrow $ $ J $) and, starting from any state, satisfies the invariant $ J $.\cite{key}\\

For differential equations, there are two possibilities or two possible paths. If the ODE has a polynomial solution, we can replace it with a discrete assignment at each point in time $ t $. In this case, we would have a polynomial for each variable that symbolically describes the value of this variable in time. Now if there is no polynomial solution available, this will lead to a dead end. So in these cases, differential induction is applied, which is induction for differential equations showing that the possible derivative of the candidate solution in the domain of evolution points inward in the region that characterizes this ODE. \cite{key} \\

Differential induction is a natural deduction technique for differential equations. It is based on the local dynamics of the differential equations, and does not need the solutions of the differential equations, because the equations are simpler than the solutions, and furthermore, the techniques to normalize Differential equations are easier than technical techniques to find solutions. \cite{Andre} \\

Now, we have already talked about the discrete part of the system, that is why now we turn to the continuous part, and we will study a model in differential equations that models the vehicular traffic.

\subsection{Vehicular Traffic Model}
Understanding the evolution and dynamics of vehicular traffic is a highly complex problem that requires the intervention of different approaches, all with the idea of ​​being able to show how vehicular traffic behaves . 

Each of these approaches is determined by the space-time scale that we want to work on, for example if we take vehicles as individuals we have microscopic models, which model the behavior of each individual. or vehicle, generally this approach is treated from cellular automata or car-following type models. We can also focus on the entire road, in this way we can introduce elements of fluid mechanics to study the behavior of traffic. For this approach, it is generally studied from the partial differential equations from the LWR models (Lighthill-Whitham-Richards), because they are based on fluid mechanics, variables such as the flow of vehicles appear ($ Q $ ), the density of the way ($ k $) and the average velocity ($ v $). And there are also higher order models that model traffic as the cinetic behavior of a gas, among others. \cite{ModTraf, Jin, Argen} \\

In this work we will focus on a general vision of the model, so we will work with ODE. This is why we will use the definition of flow ($ Q $) which is based on the fundamental triangular diagram of LWR models.
The fundamental triangular diagram defines flow as a function that depends on the state of traffic and that only depends on the density of vehicles. This flow ($ Q $) is characterized by:

\begin{equation}
	Q(\rho) =
	\left\{
	\begin{matrix}
		V_0\rho & \text{if } \rho \leq \rho_c = \dfrac{1}{V_0T + l} & \text{ (Uncongested traffic) }\\
		\dfrac{1}{T}(1-\rho l) & \text{if } \rho_c < \rho \leq \rho_{\max} = 1/l & \text{ (Congested traffic) }
	\end{matrix}
	\right.
\end{equation}
where $ V_0 $ is the desired speed, $ l $ is the average size of the cars, and $ T $ is the average time between cars. These parameters depend on the way. \\

Thanks to this characterization, the problem is simplified enough not to work with hyperbolic partial differential equations. \cite{ModTraf} This is because normally for LWR models the flow is a function that depends on density ($ \rho $) and velocity ($ v $), that is, $ Q (\rho, v) $. \\

Now, since we want to work with a road network, we will use the \cite{Jin} model:

\begin{equation}
	\dfrac{dk_a(t)}{dt}=\dfrac{1}{L_a}(f_a - g_a)
\end{equation}

Where $ k_a (t) $ corresponds to the vehicular density in each section of a network, $ f_a $ is the inflow, $ g_a $ is the outflow and $ L_a $ is the length of each section. \\

Also, since we start from an LWR model, we have the assumption that the flow only depends on the density, that is, $$ Q (t) = Q (k (t)) $$.

Now, considering the ideas of the block transmission model presented in \cite{Dagan, Lebac}, then we have two new variables that depend on the flow, which are demand (outflow) and supply (inflow). ), which are defined as:
\begin{equation}
	d_a(t)=Q_a(min\{k_a(t),k_{a,c}(t)\}) =
	\left\{
	\begin{matrix}
		Q_a(k_a(t)) & \text{si } k_a(t) \in [0,k_{a,c}] \\
		C_a & \text{si } k_a(t) \in [k_{a,c},k_{a,j}] 
	\end{matrix}
	\right.
\end{equation}

\begin{equation}
	s_a(t)=Q_a(max\{k_a(t),k_{a,c}(t)\}) =
	\left\{
	\begin{matrix}
		C_a  & \text{si } k_a(t) \in [0,k_{a,c}] \\
		Q_a(k_a(t)) & \text{si } k_a(t) \in [k_{a,c},k_{a,j}] 
	\end{matrix}
	\right.
\end{equation}

Where $ C_a $ is the capacity of the section and $ k_ {a, c}, k_ {a, j} $ are the critical density and the bottleneck density, respectively. \\

These demand and supply functions have a similar sense to the economic one, because demand is the input of products or services. For traffic, the demand is also in an input, only not of products but of flow, that is why for a given density there is a flow $ Q $ and when the density is at the maximum limit it is demand changes to $ C_a $ or the maximum capacity of the path. The same happens with the offer, since it can be seen as the output of products or services. Which for our case would be the flow output. \\
In addition to having these demand and supply functions, at work \cite{Argen}
we can find other functions associated with the types of intersections that can be found in a way.

\begin{figure}[H]
	\centering
	\includegraphics[width=1\linewidth]{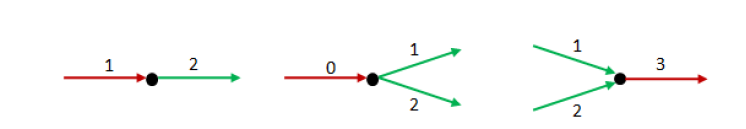}
	\caption{Types of Intersection: Lineal, Divergence, Meeting ,Source : \cite{Argen}}
	\label{fig:tdi}
\end{figure}

For the linear union we have that the input flow ($ f_a $) and the output flow ($ g_a $) is like this:

\begin{equation}
	g_1(t) = f_2(t) = \pi(t)min\{d_1(t),s2(t)\}
\end{equation}

For the divergence-type intersection, the inflows and outflows are given as follows:

\begin{equation}
	g_0(t) = \pi(t)min\{d_0(t),\frac{s_1(t)}{\xi_{0 \rightarrow 1}(t)},\frac{s_2(t)}{\xi_{0 \rightarrow 2}(t)}\}
\end{equation}

\begin{equation}	
	f_1(t) = \xi_{0 \rightarrow 1}(t)g_0(t)
\end{equation}

\begin{equation}	
	f_2(t) = \xi_{0 \rightarrow 2}(t)g_0(t)
\end{equation}

For this intersection, the function $ \xi (t) $ is the probability that the cars will turn along that road. And finally, we have the last type of intersection where the output flows and the input flow is given as follows:

\begin{equation}
	f_3(t) = min\{d_1(t)+d_2(t), s_3(t) \}
\end{equation}

\begin{equation}
	g_1(t) = min\{d_1(t),max\{s_3(t)-d_2(t),\frac{C1}{C1+C2}s_3(t)\}\}
\end{equation}

\begin{equation}	
	g_2(t) = f_3(t)-g_1(t)
\end{equation}

The function $ \pi (t) $ that appears at the intersections of linear union and divergence, represents the behavior of the traffic capacity located in that lane, and is characterized as follows:

\begin{equation}
	\pi(t) =
	\left\{
	\begin{matrix}
		1 & \text{ Sem\'aforo en verde }\\
		0 & \text{Sem\'aforo en rojo }
	\end{matrix}
	\right.
\end{equation}

\section{Hybrid Systems Implementation}
Continuing with the previously proposed model, a new part is added, which consists of taking into account the bus stops in some routes, for which a new function is built.

\begin{equation}
	P(t) =
	\left\{
	\begin{matrix}
		\psi & \text{ No Empty }\\
		1 & \text{Empty}
	\end{matrix}
	\right.
\end{equation} 
As we want to characterize the behavior of the traffic, when these stops appear on certain roads, then what we do is suppose that the behavior of these stops function as a pseudo traffic light. This can be noticed when there is no bus at the stop, the flow of traffic does not change, as it would when the traffic light is green. Now, when the stop is busy, the flow will not stop as it happens when the traffic light is red, but it will slow down in a certain proportion, that is, the flow will not stop but will be paused, since the cars will brake and they will change lanes, then we can assume that this behavior will be determined by a variable $ \psi $, which will determine how long the flow will be paused. This is why $ \psi $ is a variable that changes its value, in a probabilistic way or with some randomness, for this reason the variable is in the interval (0,1).

\subsection{Hybrid Automaton}
Now, to build the hybrid automaton, we set out to transform the model presented to a system by events, thus taking four events. Which are the types of intersection (linear union, divergence and meeting) and the bus stop. These events occur in the following way: in event one there are two lanes connected by a traffic light, in the second event there is one lane that joins two others, and in the third event there are two lanes that meet. to a single lane. For the event of the bus stop, there is also a perspective of two linked lanes, only that in one of them the bus stop is presented that will function as a pseudo traffic capacity depending on the existence of the bus.

The automaton created from the events mentioned are as follows:

\begin{figure}[H]
	\begin{subfigure}{.5\textwidth}
		\centering
		\includegraphics[width=1\linewidth]{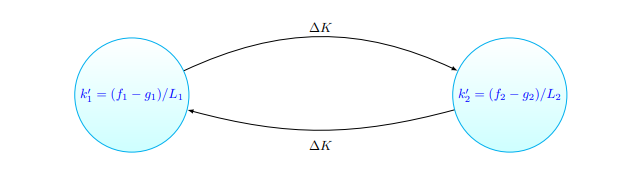}
		\caption{Lineal Union Traffic Light Hybrid Automaton}
		\label{fig:ahls}
	\end{subfigure}
	\begin{subfigure}{.5\textwidth}
		\centering
		\includegraphics[width=1\linewidth]{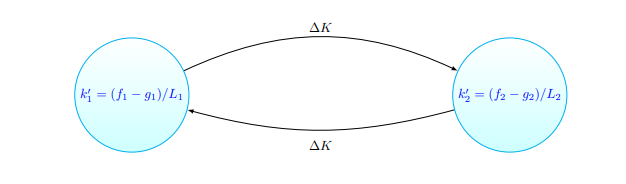}
		\caption{Lineal Union Bus Stop Hybrid Automaton}
		\label{fig:ahlp}
	\end{subfigure}
	
	\begin{subfigure}{.5\textwidth}
		\centering
		\includegraphics[width=1\linewidth]{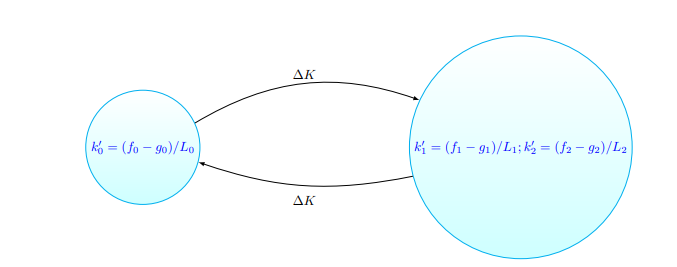}
		\caption{Divergence Hybrid Automaton}
		\label{fig:ahd}
	\end{subfigure}
	\begin{subfigure}{.5\textwidth}
		\centering
		\includegraphics[width=1\linewidth]{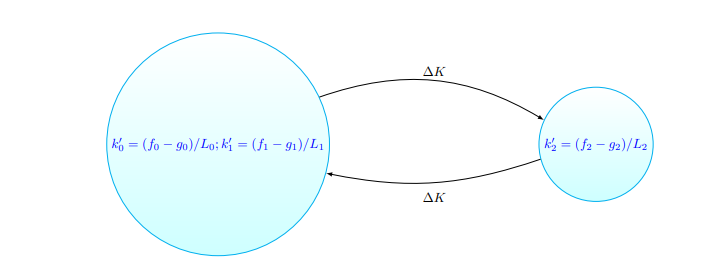}
		\caption{Meeting Hybrid Automaton}
		\label{fig:ahe}
	\end{subfigure}
	\caption{Source: Own Elaboration}
\end{figure}

The first image is that of the linear union automaton, this is made from the first event presented in the model. It can be seen that it starts from a system of differential equations that describes the behavior in a lane and because the traffic light takes one of the two possibilities (green or red), a discrete change in density occurs, thus causing a new density in the lane, and this causes the lane that follows it to have a similar behavior because the traffic capacity will allow it to have an inflow or not.

In the second image we see the bus from the bus stop. As we said earlier in this event, the existence of the bus produces a pseudo traffic capacity effect, thus producing that when the bus is at the stop, there is a change in the dynamics of both lanes, thus generating a new state and when it is not there there is also a change in density, causing it to return to another state.

In the third image is the divergence car, because there is a union of a lane with two lanes, so we can see that the car starts in a state of one lane and then when it occurs the intersection generates a change in the densities and the new generated state has two differential equations that will describe the density in each of the lanes; It should be noted that each inflow will depend on a variable of probability of turning, that is, how likely it is to continue straight or turn to the right.

And finally, we have the meeting car, for this case we start from two lanes, so we have two differential equations that describe the dynamics in those lanes and when the union of these lanes occurs this causes the density to vary and generate another state on the new lane.

\subsection{Hybrid Program}
Having presented the hybrid automatons, we can show the programs generated by these automatons.

\begin{figure}[H]
	\centering
	\includegraphics[width=1\linewidth]{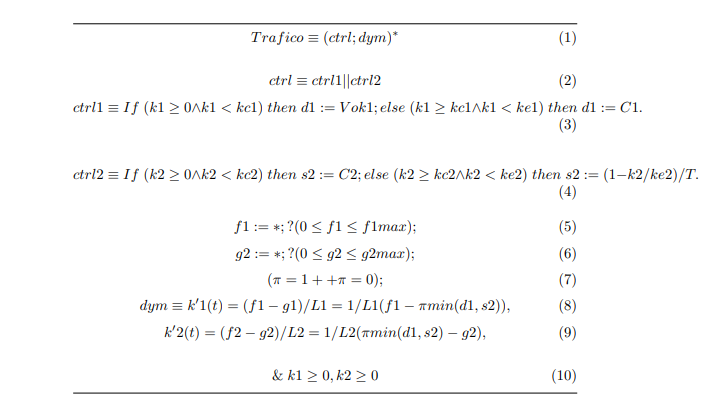}
	\caption{Hybrid Program of Lineal Union with Traffic Light}
	\label{fig:mhs}
\end{figure}
This is the program of the event of linear junction with traffic capacity, we see that the dynamics of the traffic is divided into two, one part is the control or the discrete part and the other is that of the traffic flow or the continuous part. The control depends on two controls, one for each lane, first it is evaluated to which interval the lane density belongs ($ k_a $), in order to determine the value of the demand function in lane 1 and of the function supply in lane 2. Subsequently two flows are chosen that must belong to a fixed interval, first we choose an input flow and it is determined if this chosen flow is between 0 and a maximum flow for that lane, the same is done for the output stream $ g_2 $. This choice of flows is due to the fact that in the original model an initial value must be given to the entrance flow of lane 1 and the exit of lane 2 would be another lane which will depend on the future changes that occur in that lane. In addition, the value of the traffic light ($ \pi $) is defined, the symbol ++ means that the choice of the value of $ \pi $ is non-deterministic. And finally, there is the continuous part with the differential equations for each lane and its domain of evolution.

\begin{figure}[H]
	\centering
	\includegraphics[width=1\linewidth]{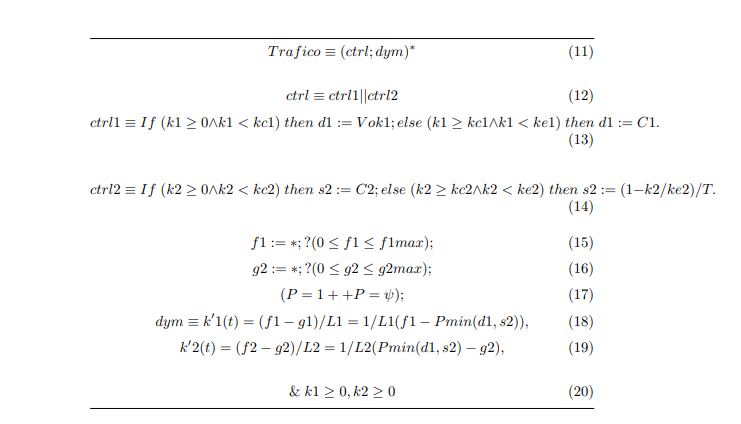}
	\caption{Hybrid Program of Lineal Union with Bus Stop}
	\label{fig:mhp}
\end{figure}
The second program does not vary much from the first, that is, we continue to have two lanes and two controls on the densities of these lanes, in addition, an input flow and an output flow are also chosen, and we have the non-deterministic assignment of P, that is, whether the bus is at the stop or not. And finally, it is at the dynamics of these lanes. This model is the same or similar to the previous one due to the fact that we can assume that the bus stop causes a dynamics similar to that of the traffic capacity.

\begin{figure}[H]
	\centering
	\includegraphics[width=1\linewidth]{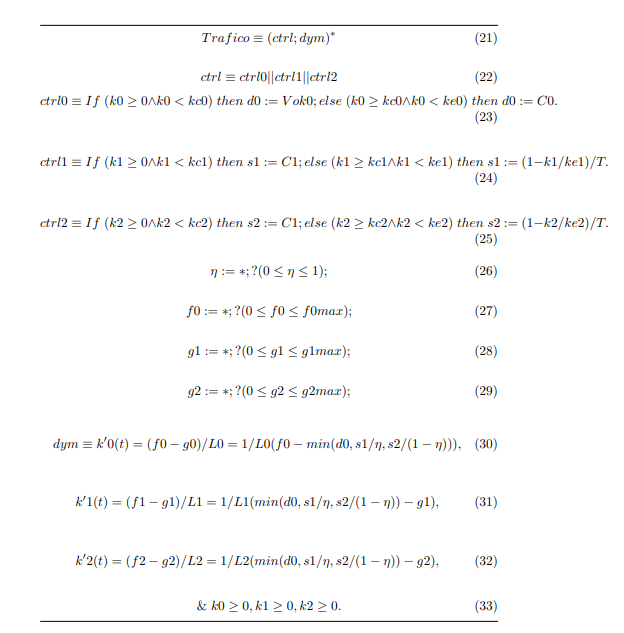}
	\caption{Hybrid Program of Divergence}
	\label{fig:mhd}
\end{figure}
This is the program of the divergence event, the interpretation of the traffic is the same, that is, there is a control over the density of each lane and there is a continuous part. In the controls, because there are 3 lanes there are 3 controls, where again in each control it is selected in which interval each density is ($ k_0, k_1, k_2 $) in order to choose the values ​​of the demand and supply functions , respectively. Now, the variable $ \eta $ is the probability variable of turning, that is, it is the variable that determines how likely it is to turn to lane 2 or to continue straight on lane 1. There is also an assignment or choice of three streams, one input and two output. And finally there are the three differential equations with their respective domains of evolution.

\begin{figure}[H]
	\centering
	\includegraphics[width=1\linewidth]{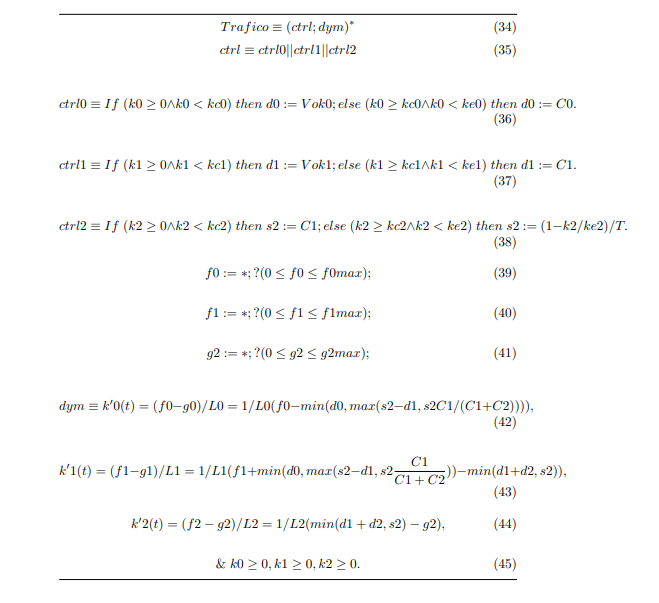}
	\caption{Hybrid Program of Meeting}
	\label{fig:mde}
\end{figure}

This is the last program associated with the meeting event, and like the previous one, it is based on an interpretation of the traffic as the union of a discrete or control part and a continuous part or dynamic. Also like the previous program, the control is divided into three parts, one for each lane present in the event. In addition, each density is selected taking into account the intervals to which they belong, and the value of the demand and supply functions is also defined. An assignment to two input streams and one output stream is also presented. And finally, there are the differential equations associated with each lane, with their respective domains of evolution.

\section{Verification}
Now, we move on to the verification of the programs associated with each event. This verification is done through the \textit{model checking} used by the KeYmaeraX program. This verification consists of using through different tactics or steps the calculation of d$L$ sequences, in order to validate the correctness of each state presented by the program associated with each event. Once the axiom has been reached, that is, it has reached empty, it can be said that the program as a whole is valid, thus producing that if all the programs are valid then we could say that the model provided in \cite{Argen} is valid, not only numerically but also logically. \\

Once the program is passed to KeYmaeraX, it shows the tactics that I used to validate the program and also shows the derivation tree, as seen in the following figure.

\begin{figure}[H]
	\begin{subfigure}{.5\textwidth}
		\centering
		\includegraphics[width=1.3\linewidth]{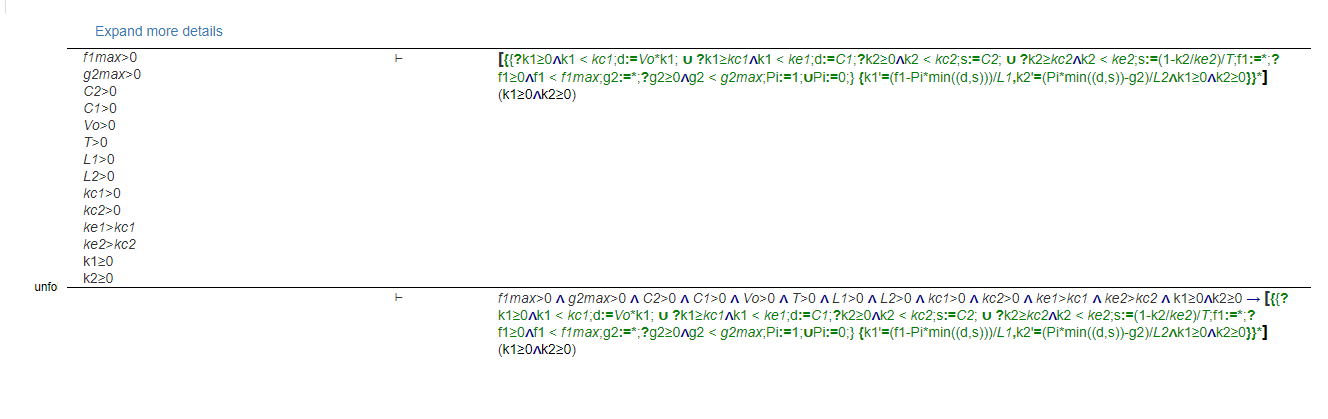}
		\caption{Proof Tree of Lineal Union with Traffic Light}
		\label{fig:ptls}
	\end{subfigure}
	\begin{subfigure}{.5\textwidth}
		\centering
		\includegraphics[width=1.3\linewidth]{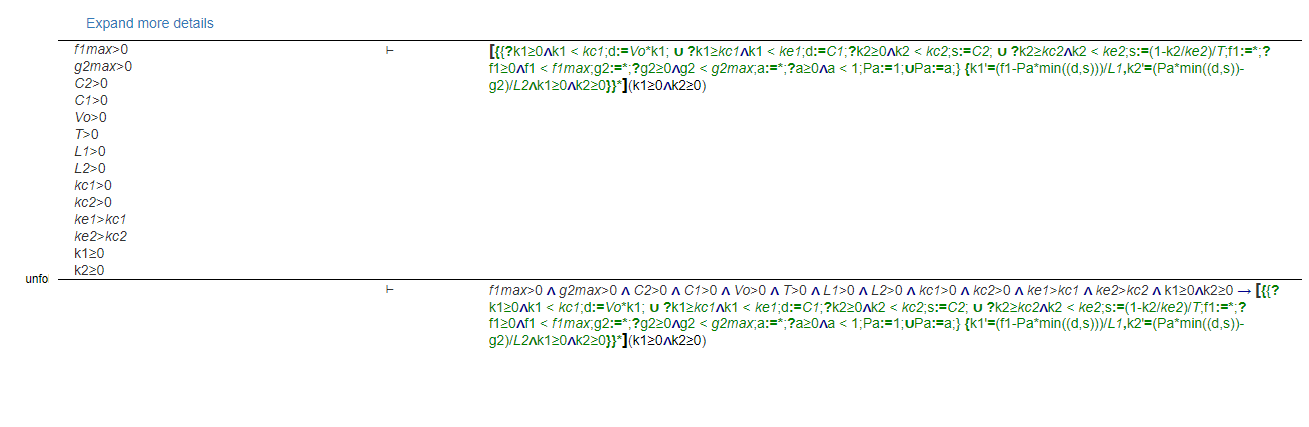}
		\caption{Proof Tree of Lineal Union with Bus Stop}
		\label{fig:ptlp}
	\end{subfigure}
	
	\begin{subfigure}{.5\textwidth}
		\centering
		\includegraphics[width=1.3\linewidth]{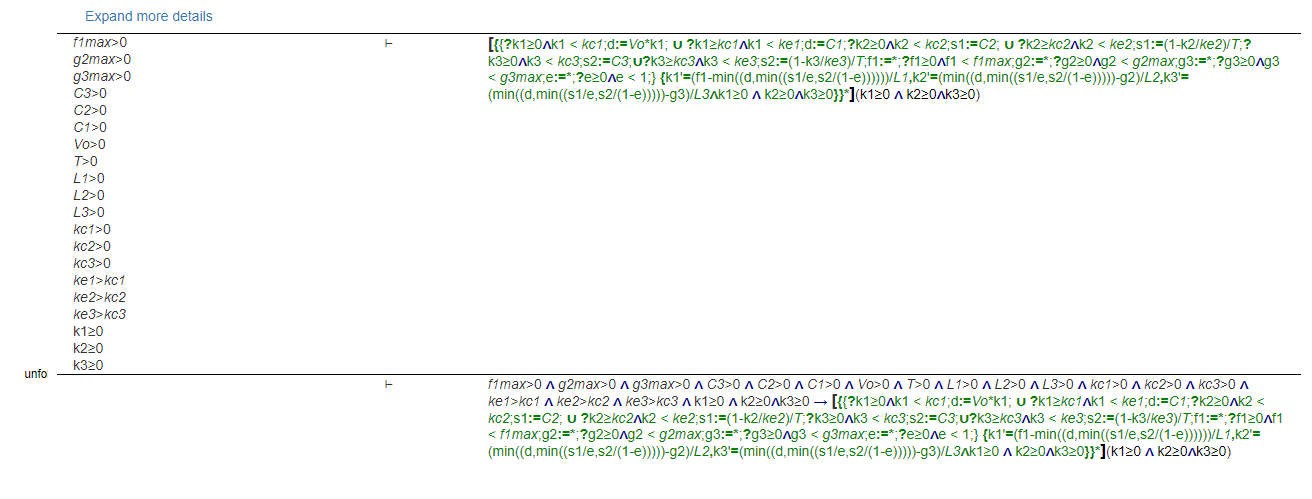}
		\caption{Proof Tree of Divergence}
		\label{fig:ptd}
	\end{subfigure}
	\begin{subfigure}{.5\textwidth}
		\centering
		\includegraphics[width=1.3\linewidth]{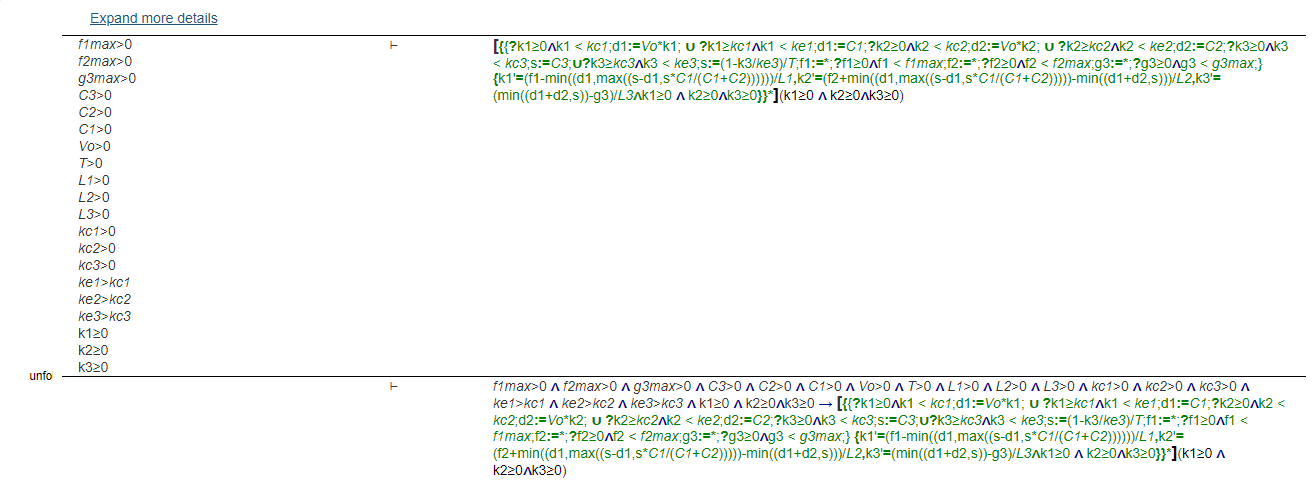}
		\caption{Proof Tree of Meeting}
		\label{fig:pte}
	\end{subfigure}
\end{figure}
As we see the result of the verification of KeYmaeraX, it is that each flattened tree has reached the axiom. But to better detail the tree and understand what KeYmaeraX did, we will take the first tree separately, and it will be arranged in a non-flat way so that we can analyze the flat and the non-flat tree.

\begin{figure}[H]
	\centering
	\includegraphics[width=1.1\linewidth]{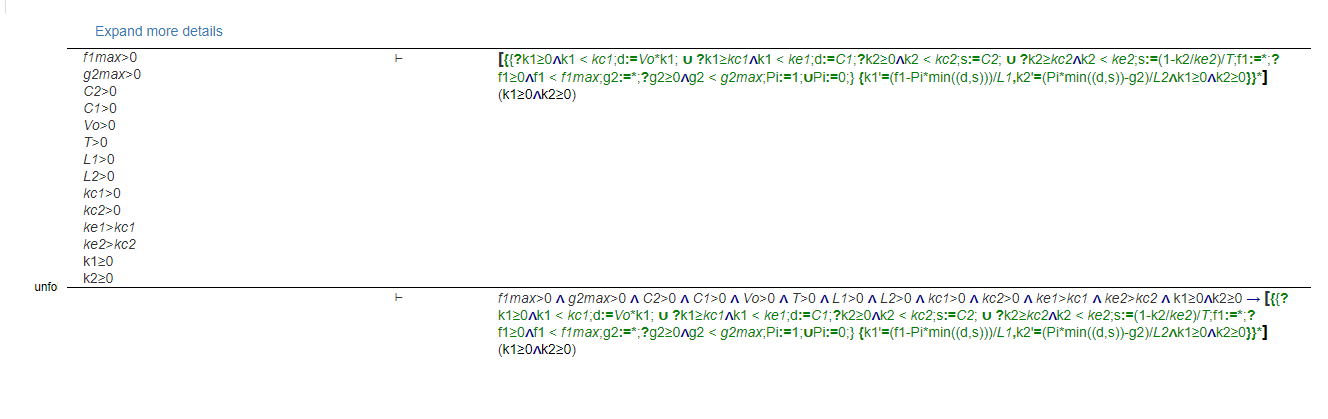}
	\caption{Proof Tree of Lineal Union with Traffic Light}
	\label{fig:ptls2}
\end{figure}
\vspace{2cm}

\begin{prooftree}
	\AxiomC{}
	\LeftLabel{master tactic}
	\UnaryInfC{$ Hip  \vdash  (EscVar;AsgFl;AsgPi;Dym)^*$}
	\LeftLabel{$\rightarrow$ r}
	\UnaryInfC{$ \vdash Hip \rightarrow (EscVar;AsgFl;AsgPi;Dym)^*$}
\end{prooftree}
\vspace{2cm}

We take the first program, that is, the linear union program with semaphore. We see that on one side the flattened tree that KeYmaeraX delivers and on the other we see the normal derivation tree. For the non-flattened tree, the conditions were given names so that they would fit in the document. In the following table we see the association of each name with a condition.

\begin{table}[H]
	\begin{adjustbox}{width=1\linewidth}
		\begin{tabular}{|l|l|}
			\hline
			Hip    & $f1max>0, g2max>0, C2>0, C1>0, Vo>0, T>0, L1>0, L2>0, kc1>0, kc2>0, ke1>kc1, ke2>kc2, k1\geq 0, k2\geq 0$                                                      \\ \hline
			EscDen & $?(k1\geq 0 \wedge k1<kc1);d:=Vo*k1; ++ ?(k1\geq kc1 \wedge k1<ke1);d :=C1; ?(k2\geq 0 \wedge k2<kc2); s:=C2; ++ ?(k2\geq kc2 \wedge k2<ke2); s:=(1-k2/ke2)/T$ \\ \hline
			AsgFl  & $f1:=;?(f1\geq 0 \wedge f1<f1max); g2:=;?(g2\geq 0 \wedge g2<g2max)$                                                                                         \\ \hline
			AspPi  & $\pi :=1; ++ \pi :=0$                                                                                                                                          \\ \hline
			Dym    & $k1'=(f1-Pi*min(d,s))/L1, k2'=(Pi*min(d,s)-g2)/L2 \wedge (k1\geq 0 \wedge k2\geq 0)$                                                                           \\ \hline
		\end{tabular}
	\end{adjustbox}
\end{table}

Therefore, Hip are the hypotheses, EscDen are the controls over the densities, AsgFl is the allocation of the flows, AsigPi is the allocation of the $ \pi $ function, and Dym is the system of differential equations for each lane. Now, we see that KeYmaeraX uses a tactic called \"master tactic". This tactic is used to verify all the content of the program and arrive at the axiom. This tactic is KeYmaeraX's automatic way of solving problems. This tactic is actually a set of different types of tactics created by KeYmaeraX developers, with which the program executes the tactic that it thinks will solve the problem, this set of tactics is always getting bigger, that is, the developers They are always adding more tactics so that KeYmaeraX can face any type of hybrid system that comes its way. \cite{MasTac}

\section{Conclusions}

\begin{itemize}
	\item It was possible to determine a model of ordinary differential equations that would successfully model vehicular traffic.
	
	\item The representation could be built through hybrid systems of the chosen model, that is, the automatons that correspond to each of the 3 particular events described by the model were built and they were I add a fourth event. The models associated with these automaton were also built. 
	
	\item The KeYmaeraX program was used to verify the correctness of the system, in this case, the different programs obtained, and the validity of each of the programs was achieved.
	
\end{itemize}

\nocite{*}
\bibliographystyle{fundam}


\end{document}